\documentclass[iop,apj]{emulateapj}

\usepackage{epsfig}
\usepackage{graphicx,rotating,ragged2e}
\usepackage{amsmath}

\begin{document}

%===================================================
% Title and Author Info
%===================================================
\title{\lowercase{i}PTF14\lowercase{yb}: The First 
Discovery of a GRB Afterglow Independent of a High-Energy
Trigger}

\author{
S.~Bradley Cenko\altaffilmark{1,2},
  Alex L.~Urban\altaffilmark{3},
  Daniel A.~Perley\altaffilmark{4,5},
  Assaf Horesh\altaffilmark{6},  
  Alessandra Corsi\altaffilmark{7},  
  Derek B.~Fox\altaffilmark{8},  
  Yi Cao\altaffilmark{4},
  Mansi M.~Kasliwal\altaffilmark{9},
  Amy Lien\altaffilmark{1,10},
  %% Alphabetical from here on out
  Iair Arcavi\altaffilmark{11,12},
  Joshua S.~Bloom\altaffilmark{13},
  Nat R.~Butler\altaffilmark{14,15},
  Antonino Cucchiara\altaffilmark{1},
  Jos\'{e} A.~de Diego\altaffilmark{16},
  Alexei V.~Filippenko\altaffilmark{13},
  Avishay Gal-Yam\altaffilmark{6},
  Neil Gehrels\altaffilmark{1},
  Leonid Georgiev\altaffilmark{15},
  J.~Jes\'{u}s Gonz\'{a}lez\altaffilmark{15},
  John F.~Graham\altaffilmark{17},
  Jochen Greiner\altaffilmark{17},
  D.~Alexander Kann\altaffilmark{18},
  Christopher R.~Klein\altaffilmark{13},
  Fabian Knust\altaffilmark{17},
  S.~R.~Kulkarni\altaffilmark{4},
  Alexander Kutyrev\altaffilmark{1},
  Russ Laher\altaffilmark{19},
  William H.~Lee\altaffilmark{15},
  Peter E.~Nugent\altaffilmark{20,16},
  J.~Xavier~Prochaska\altaffilmark{21},
  Enrico Ramirez-Ruiz\altaffilmark{21},
  Michael G.~Richer\altaffilmark{15},
  Adam Rubin\altaffilmark{6},
  Yuji Urata\altaffilmark{22},
  Karla Varela\altaffilmark{17},
  Alan M.~Watson\altaffilmark{15},
  and Przemek R.~Wozniak\altaffilmark{23}
}

\altaffiltext{1}{Astrophysics Science Division, NASA Goddard Space Flight Center,
Mail Code 661, Greenbelt, MD 20771, USA}
\altaffiltext{2}{Joint Space-Science Institute, University of Maryland, College 
Park, MD 20742, USA}
\altaffiltext{3}{Leonard E. Parker Center for Gravitation, Cosmology and Astrophysics,
  University of Wisconsin-Milwaukee, Milwaukee, WI 53211, USA}
\altaffiltext{4}{Astronomy Department, California Institute of 
  Technology, Pasadena, CA 91125, USA}
\altaffiltext{5}{Hubble Fellow}
\altaffiltext{6}{Benoziyo Center for Astrophysics, Weizmann Institute of Science,
   76100 Rehovot, Israel}
\altaffiltext{7}{Department of Physics, Texas Tech University, Box 41051,
   Lubbock, TX 79409-1051, USA}
\altaffiltext{8}{Department of Astronomy \& Astrophysics, Pennsylvania State 
   University, University Park, PA 16802, USA}
\altaffiltext{9}{Observatories of the Carnegie Institution for Science, 813
   Santa Barbara St., Pasadena, CA, 91101, USA}
\altaffiltext{10}{Department of Physics, University of Maryland, Baltimore 
   County, Baltimore, MD 21250, USA}
\altaffiltext{11}{Las Cumbres Observatory Global Telescope, 6740 Cortona Dr, 
   Suite 102, Goleta, CA 93111, USA}
\altaffiltext{12}{Kavli Institute for Theoretical Physics, University of 
   California, Santa Barbara, CA 93106, USA}
\altaffiltext{13}{Department of Astronomy, University of California, Berkeley, 
   CA 94720-3411, USA}
\altaffiltext{14}{School of Earth and Space Exploration, Arizona State 
   University, Tempe, AZ 85287, USA}
\altaffiltext{15}{Cosmology Initiative, Arizona State University, Tempe, 
   AZ 85287, USA}
\altaffiltext{16}{Instituto de Astronom\'{i}a, Universidad Nacional 
   Aut\'{o}noma de M\'{e}xico, Apartado Postal 70-264, 04510 M\'{e}xico, 
   D.~F., M\'{e}xico}
\altaffiltext{17}{Max-Planck-Institut f{\"u}r extraterrestrische Physik,
   Giessenbachstra{\ss}e 1, 85748, Garching, Germany}
\altaffiltext{18}{Th{\"u}ringer Landessternwarte Tautenburg, Sternwarte 5, 
   07778 Tautenburg, Germany}
\altaffiltext{19}{Spitzer Science Center, California Institute of Technology, 
   M/S 314-6, Pasadena, CA 91125, USA}
\altaffiltext{20}{Computational Cosmology Center, Lawrence Berkeley National 
   Laboratory, 1 Cyclotron Road, Berkeley, CA 94720, USA}   
\altaffiltext{21}{Department of Astronomy and Astrophysics and UCO/Lick 
   Observatory, University of California, Santa Cruz, CA 95064, USA}
\altaffiltext{22}{Institute of Astronomy, National Central University, 
   Chung-Li 32054, Taiwan}
\altaffiltext{23}{Los Alamos National Laboratory, Los Alamos, New Mexico 
   87545, USA}
   
\email{Email: brad.cenko@nasa.gov}

%===================================================
% Other Misc Info
%===================================================
%\slugcomment{}

\shorttitle{The First Optically Discovered GRB}
\shortauthors{Cenko et al.}

%===================================================
%===================================================

%%%%%%%%%%%%%%%%%%%%%%%%%%%%%%%%%%%%%%%%%%%%%%%%%%%%%%%%%%
% New Commands
%%%%%%%%%%%%%%%%%%%%%%%%%%%%%%%%%%%%%%%%%%%%%%%%%%%%%%%%%%
\newcommand{\Swift}{\textit{Swift}}
\newcommand{\fermi}{\textit{Fermi}}
\newcommand{\mgii}{\ion{Mg}{2} $\lambda \lambda$ 2796, 2803}
\newcommand{\oii}{[\ion{O}{2} $\lambda$ 3727]}
\newcommand{\ip}{\textit{i$^{\prime}$}}
\newcommand{\zp}{\textit{z$^{\prime}$}}
\newcommand{\gp}{\textit{g$^{\prime}$}}
\newcommand{\rp}{\textit{r$^{\prime}$}}
\newcommand{\up}{\textit{u$^{\prime}$}}
%===================================================

%===================================================
% Abstract
%===================================================

\begin{abstract}
We report here the discovery by the Intermediate Palomar Transient 
Factory (iPTF) of iPTF14yb, a luminous ($M_{r}\approx-27.8$\,mag),
cosmological (redshift 1.9733), rapidly fading optical transient.  We 
demonstrate, based on probabilistic arguments and a comparison with
the broader population, that iPTF14yb is the optical afterglow
of the long-duration gamma-ray burst GRB\,140226A.  This marks 
the first unambiguous discovery of a GRB afterglow prior to (and
thus entirely independent of) an associated high-energy trigger.
We estimate the rate of iPTF14yb-like sources (i.e., 
cosmologically distant relativistic explosions) based on iPTF
observations, inferring an all-sky value of 
$\Re_{\mathrm{rel}}=610$\,yr$^{-1}$ (68\% confidence interval of 
110--2000\,yr$^{-1}$).
Our derived rate is consistent (within the large uncertainty) with
the all-sky rate of on-axis GRBs derived by the \textit{Swift} satellite.
Finally, we briefly discuss the implications
of the nondetection to date of {\it bona fide} ``orphan'' afterglows
(i.e., those lacking detectable high-energy emission) on
GRB beaming and the degree of baryon loading in these
relativistic jets.
\end{abstract}

%===================================================
% Keywords
%===================================================

%===================================================
% Introduction
%===================================================
\section{Introduction}
\label{sec:intro}
Two central tenets of our standard model  of 
long-duration gamma-ray bursts (GRBs) hold that these explosions are 
ultrarelativistic (initial Lorentz factor $\Gamma_{0} \gtrsim 100$) and 
highly collimated (biconical jets with half-opening angle $\theta 
\approx1$--10$^{\circ}$).  The former is invoked to explain the so-called 
``compactness'' problem: absent this ultrarelativistic 
expansion, the ejecta would be optically thick to pair production at 
typical peak spectral energies of a few hundred keV, whereas the prompt 
emission is observed to be 
nonthermal\footnote{More recently, a possible photospheric component
has been identified in the prompt high-energy spectra of a number
of GRBs (e.g., \citealt{raz+09,gcb+11}), but this does not dramatically
ease the requirement of ultrarelativistic expansion.}.  On the 
other hand, a high degree of collimation is required for basic energy 
conservation: the isotropic energy release can in some cases exceed 
$10^{54}$\,erg, comparable to the rest-mass energy of their massive-star 
progenitors.

In order to accelerate material to these velocities, the outgoing jet must
entrain a very small amount of mass 
($M_{\mathrm{ej}}\approx10^{-5}$\,M$_{\odot}$); this is referred to as 
the ``baryon loading'' of
the jet.  Most \textit{observed} GRB prompt spectra,
with peak spectral energies of a few hundred keV, therefore indicate 
very ``clean'' outflows (i.e., low mass of entrained baryons; \citealt{mr92}).  
But there is growing evidence that the intrinsic population of
long GRBs is dominated by bursts with peak energies below the traditional
$\gamma$-ray bandpass (e.g., \citealt{rgk+05,bbp10}).  Could these lower 
$E_{\mathrm{pk}}$, fainter outbursts (e.g., X-ray flashes; \citealt{hzk+01})
result from an outflow with more entrained mass (i.e., a ``dirty''
fireball; \citealt{dcm00,hdl02})?  Or can other properties, such as viewing
angle \citep{grp05} or the nature of the remnant \citep{mdn+06}
account for these softer events?

Separately, the high degree of collimation requires that most ($f_{b} 
\equiv (1 - \cos{\langle \theta \rangle})^{-1}\approx100$; \citealt{gd07}) 
GRBs are in fact beamed away from us on Earth.  The afterglows of these 
off-axis bursts become visible at late times ($t \gg \Delta t_{\mathrm{GRB}}$) 
when the outflow slows down and illuminates an increasing fraction of the 
sky \citep{r99,sph99}.
Yet despite concerted efforts at uncovering such orphan afterglows  in
the X-ray \citep{ghv+00,np03,lrk04}, optical \citep{bwb+04,raa+05,rgs06}, and 
radio \citep{gop+06} bandpasses, no {\it bona fide} off-axis candidate has been
identified thus far.  

%%%%%%%%%%%%%%%%%%%%%%%%%%%%%%%%%%%%%%%%%%%%%%%%%%%%%%%%%%%%
\begin{figure*}[t!]
  \centerline{\includegraphics[width=18cm]{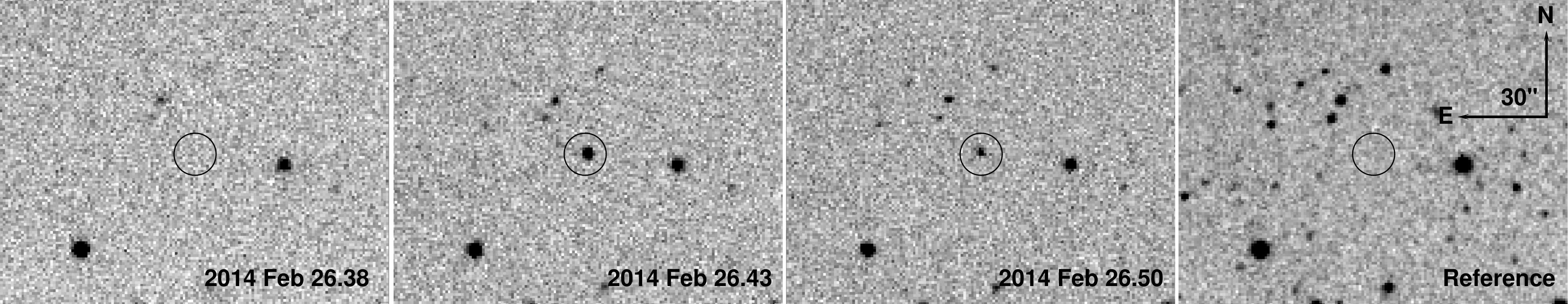}}
  \centerline{\includegraphics[width=18cm]{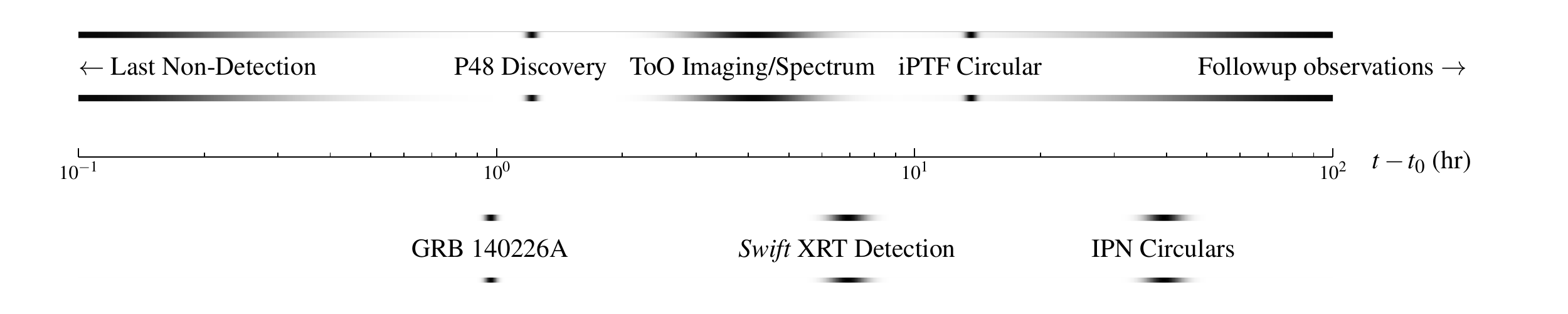}}
  \caption[]{%\small\it
  \textit{Top:} P48 discovery images of iPTF14yb.  The circle marks
  the location of iPTF14yb.  
  \textit{Bottom:} Timeline of iPTF14yb/GRB\,140226A discovery and
  announcements.}
  \label{fig:finder}
\end{figure*}
%%%%%%%%%%%%%%%%%%%%%%%%%%%%%%%%%%%%%%%%%%%%%%%%%%%%%%%%%%%

All of these issues can be addressed by sensitive, wide-field surveys that
target relativistic explosions independent of any high-energy trigger.
To that end, we present here the discovery by the 
Intermediate Palomar Transient  Factory (iPTF; \citealt{lkd+09})
of iPTF14yb, a luminous ($M_{r}\approx-27.8$\,mag), rapidly fading
optical transient at redshift $z=1.9733$.  We demonstrate that this
object is very likely associated with GRB\,140226A, making iPTF14yb the first 
unambiguous example of a GRB afterglow discovered independent of a 
high-energy trigger.

Throughout this work, we adopt a standard $\Lambda$CDM cosmology with
parameters from \citet{aaa+14b}.  All quoted uncertainties are 1$\sigma$ 
(68\%) confidence intervals unless otherwise noted, and UTC times are 
used throughout.

%===================================================
% Discovery and Observations
%===================================================
\section{Discovery and Follow-Up Observations}
\label{sec:obs}
\begin{figure*}[t!]
  \centerline{\includegraphics[width=10cm]{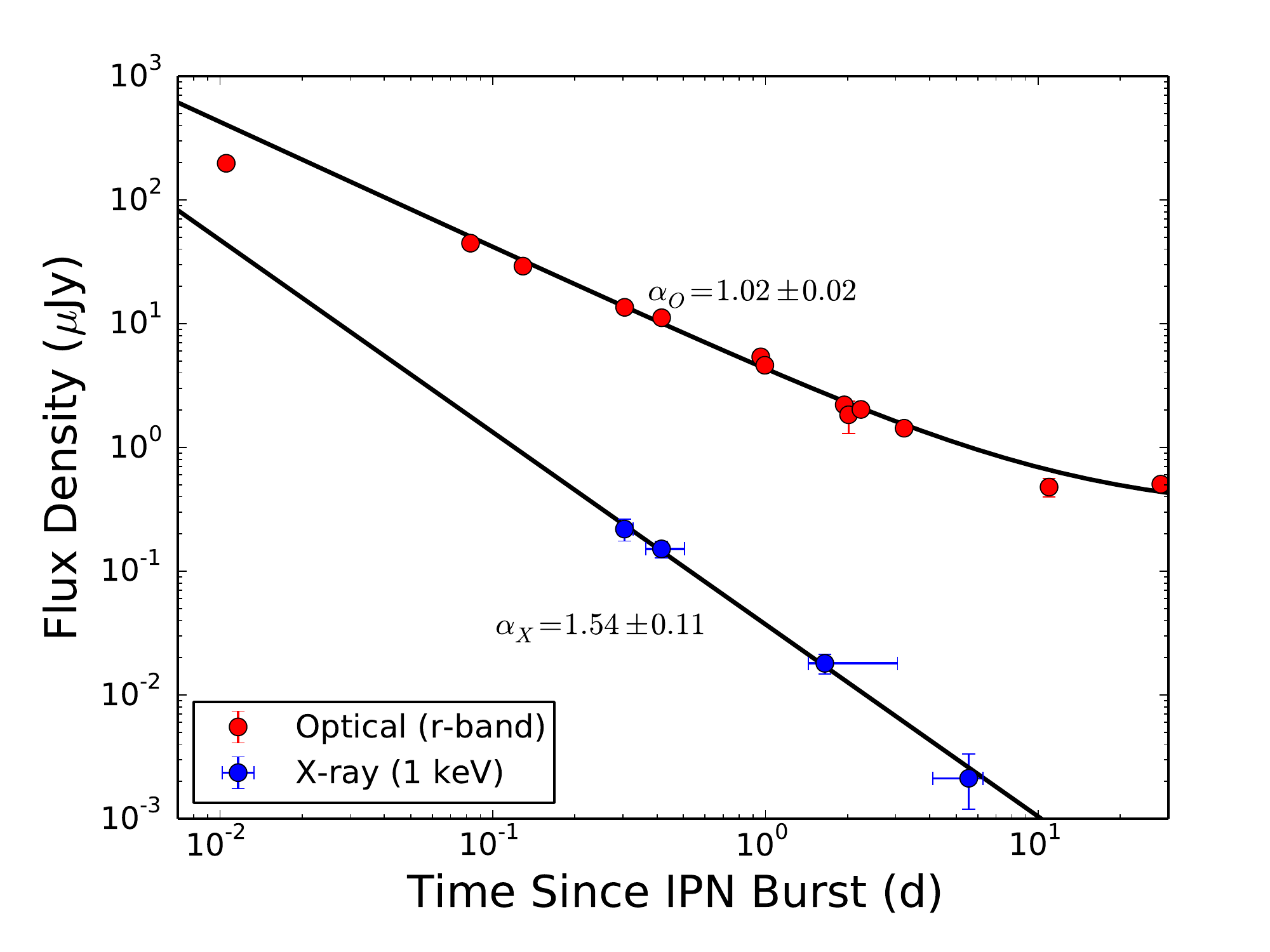}
              \includegraphics[width=9cm]{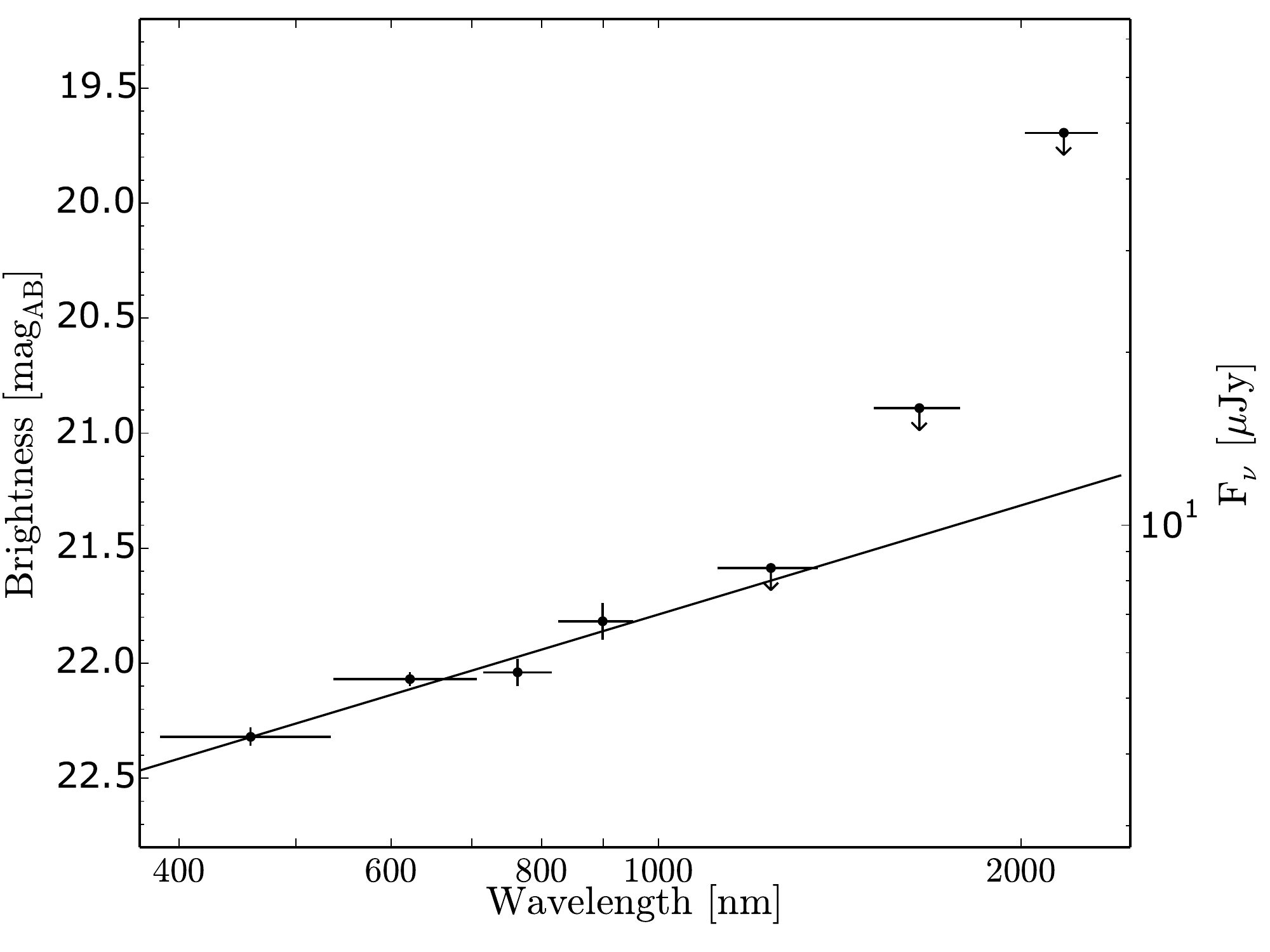}}
  \caption[]{%\small\it
  \textit{Right:} Optical ($r$-band) and X-ray light curves of iPTF14yb.  
  The outburst onset is taken as the time of the IPN GRB\,140226A.
  \textit{Left:} Optical/near-infrared SED at $\Delta t=1.0$\,d.}
  \label{fig:lcurve}
\end{figure*}

As part of regular monitoring observations with the Palomar 48\,inch
Oschin Schmidt telescope (P48)\footnote{P48 data processing is described
by \citet{lsg+14b}, while photometric calibration of iPTF data is 
discussed by \citet{oll+12}.}, we 
discovered a new transient source, designated iPTF14yb, at J2000.0 
location $\alpha=14^{\mathrm{h}}45^{\mathrm{m}}58.01^{\mathrm{s}}$, 
$\delta=+14^{\circ} 59\arcmin 35\farcs1$ (estimated uncertainty of
80 mas in each coordinate; Figure~\ref{fig:finder}).  iPTF14yb was
first detected in a 60\,s image beginning at 10:17:37 on 2014 Feb. 26,
with a magnitude of $r^{\prime}=18.16\pm0.03$.
Subsequent P48 monitoring revealed
rapid intranight fading from the source (Figure~\ref{fig:lcurve}).

Nothing was detected at the location of iPTF14yb in a P48 image
beginning at 09:04:46 on 2014 Feb. 26 (i.e., 1.21\,hr before
the first detection) to a limit of $r^{\prime} > 21.16$\,mag.  A coaddition
of all existing iPTF P48 images of this location, spanning the time range
from 2009 May 28 to 2014 Feb. 24, also reveals no quiescent
counterpart to $r^{\prime} > 22.9$\,mag (Figure~\ref{fig:finder}).

Motivated by the rapid fading and lack of a quiescent counterpart,
the duty astronomer (A.~Rubin) distributed an immediate alert to the 
collaboration.  We triggered multiwavelength follow-up observations 
at a variety of facilities.  We report here
photometry obtained with the Triple-Range Imager and POLarimeter (TRIPOL)
on the 1\,m telescope at Lulin Observatory, the Gamma-Ray Burst
Optical/Near-Infrared Detector (GROND; \citealt{gbc+08}) on the 2.2\,m
telescope at ESO La Silla, the Reionization and Transients InfraRed camera
(RATIR; \citealt{bkf+12,okr+12}) on the 1.5\,m telescope on San Pedro Martir,
the Low Resolution Imaging Spectrometer (LRIS; \citealt{occ+95}) on the
10\,m Keck I telescope, the Inamori-Magellan Areal Camera and Spectrograph
(IMACS; \citealt{dbh+11}) on the 6\,m Baade telescope at Las Campanas Observatory,
and the DEep Imaging Multi-Object Spectrograph (DEIMOS; \citealt{fpk+03})
on the 10\,m Keck II telescope.  All imaging data were 
reduced in the standard manner and calibrated with respect to nearby
point sources from the Sloan Digital Sky Survey (SDSS; \citealt{aaa+14a})
in the optical and the Two Micron All-Sky Survey (2MASS; \citealt{scs+06})
in the near-infrared.  The resulting photometry is displayed in 
Table~\ref{tab:photometry}, while the $R$/$r$-band light curve is plotted
in Figure~\ref{fig:lcurve}.

We obtained target-of-opportunity X-ray observations with the 
\textit{Swift} satellite \citep{gcg+04} beginning at 17:11 on 2014
Feb. 26.  A bright
counterpart was identified in the X-Ray Telescope (XRT; \citealt{bhn+05})
images at the location of iPTF14yb.  The
resulting light curve, processed with the automated GRB analysis 
tools of \citet{ebp+09}, is plotted in Figure~\ref{fig:lcurve}.
The X-ray spectrum is well described by a power law with a photon
index $\Gamma=2.1^{+0.5}_{-0.3}$.

The position of iPTF14yb was observed with the Karl G. Jansky Very 
Large Array (VLA\footnote{The National Radio Astronomy Observatory 
is a facility of the National Science Foundation (NSF) operated under 
cooperative agreement by Associated Universities, Inc.}) in its A 
configuration under program 14A-483 (PI~S. 
Kulkarni).  Two epochs were obtained, one on 2014 Feb. 27.7 (C- and
K-bands), and one on 2014 March 25.4 (C-band).  Both observations were
conducted with the standard wide-band continuum imaging setup.
We used 3C~286 and J1446+1721 for flux and phase calibration, 
respectively.  No radio emission was detected from iPTF14yb to the 
following 3$\sigma$ limits: $f_{\nu}$(6 GHz) $< 45$\,$\mu$Jy 
(Feb. 27); $f_{\nu}$(22 GHz) $< 153$\,$\mu$Jy (Feb. 27); $f_{\nu}$(6 GHz) 
$< 16$\,$\mu$Jy.  Similarly, observations with the 
Combined Array for Research in Millimeter Astronomy
(CARMA) on 2014 Feb. 28.5 failed to detect any emission at 95\,GHz
to a 3$\sigma$ limit of $< 0.75$\,mJy.

Finally, we obtained a CCD spectrum of iPTF14yb with LRIS beginning at 
15:26 on 2014 Feb. 26.
The instrument was configured with the 400/8500 grating on the red arm,
the 600/4000 grism on the blue arm, the 560 dichroic beamsplitter, and
a 1\arcsec\-wide slit.  As a result, our spectrum provides continuous coverage 
from the atmospheric cutoff ($\lambda\approx3250$\,\AA) to 10,200\,\AA, 
with a spectral resolution of 7.1 (4.0) \AA\ on the red (blue) arm.  
%The 2D spectrum was optimally extracted, and dispersion-corrected
%using spectra of calibrated arcs.  Flux calibration was performed using
%spectra of the standard star BD+262606 taken later that evening.  
The resulting one-dimensional spectrum is plotted in Figure~\ref{fig:spec1}.

Superimposed on a relatively flat continuum ($f_{\lambda} \propto 
\lambda^{-1.3 \pm 0.1}$), we identify strong metal absorption lines
from \ion{Mg}{2}, \ion{Fe}{2}, \ion{Al}{2}, \ion{C}{4}, \ion{Si}{2},
\ion{Si}{4}, \ion{C}{2}, and \ion{O}{1} at 
$z=1.9733\pm0.0003$.  A damped Lyman $\alpha$ (DLA) system
with $\log(N_{\mathrm{H\,I}}/{\rm cm}^2)=20.7\pm0.2$ is also observed
at this redshift, and the onset of the Lyman $\alpha$ forest blueward
of \ion{H}{1} implies that this is the redshift of iPTF14yb.

%%%%%%%%%%%%%%%%%%%%%%%%%%%%%%%%%%%%%%%%%%%%%%%%%%%%%%%%%%%%
%%%%%%%%%%%%%%%%%%%%%%%%%%%%%%%%%%%%%%%%%%%%%%%%%%%%%%%%%%%

%===================================================
% Association with GRB\,140226A
%===================================================
\section{Association with GRB\,140226A}
\label{sec:grb}
Following notification of our discovery of iPTF14yb, the Inter-Planetary Network of
high-energy detectors (IPN: \citealt{hga+10}) reported the discovery of 
GRB\,140226A, a possible counterpart of iPTF14yb \citep{GCN.15888}.
GRB\,140226A was detected by the \textit{Odyssey}, \textit{INTEGRAL}, and
\textit{Konus} satellites at 10:02:57 on 2014 Feb. 26; this is 14.7\,min 
before the midpoint of our P48 discovery image, and 58.2\,min after our 
last P48 nondetection.  The \textit{Konus} light curve{\footnote{See \url{http://www.ioffe.rssi.ru/LEA/GRBs/GRB140226A}.}
 shows a single pulse with a duration
of 15\,s (i.e., a long-duration GRB), and a 20\,keV -- 10\,MeV 
$\gamma$-ray fluence of $(5.6\pm1.1)\times10^{-6}$\,erg\,cm$^{-2}$
\citep{GCN.15889}.  
%The time-averaged spectrum is well fit by a cut-off
%exponential model with $\alpha = -1.1 \pm 0.1$ and $E_{\mathrm{pk}} =
%414 \pm 79$\,keV \citep{GCN.15889}.  
At this time, the location of 
iPTF14yb was below the horizon for the Burst Alert
Telescope (BAT) onboard \textit{Swift}, while the 
Gamma-Ray Burst Monitor (GBM) on \textit{Fermi}
was turned off owing to passage through the South Atlantic Anomaly
\citep{GCN.15888}.

We can estimate the \textit{a posteriori} probability of chance 
coincidence, both spatially
and temporally.  The IPN localized GRB\,140226A to an annulus with an 
area of 210\,deg$^{2}$ \citep{GCN.15888}.  Thus, the likelihood of chance 
spatial association is $\sim0.005$.  Similarly, since 2010 Jan. 1,
the IPN has been detecting GRBs at a rate of $\sim0.88$\,d$^{-1}$.  
Therefore, the likelihood of an unrelated IPN GRB being detected within the 
73\,min period between the last P48 upper limit and the first detection 
of iPTF14yb is $\sim0.044$.  Hence,
the joint probability of chance coincidence is quite small, 
$\sim2\times10^{-4}$.  We conclude that iPTF14yb is very likely associated
with GRB\,140226A and shall proceed with this assumption for the
remainder of this work.

%===================================================
% iPTF14yb in the Long GRB Context
%===================================================
\section{iPTF14yb in the Long-Duration GRB Context}
\label{sec:context}
We now compare the observed properties of iPTF14yb and its host galaxy
with the known population of long-duration GRBs as a final consistency check.
We fit the X-ray light curve to a power law of the form $f_{\nu} \propto
t^{-\alpha}$, finding $\alpha_{\rm X}=1.54\pm0.11$ ($\chi^{2}=0.46$ for 2
degrees of freedom).  
At late times ($\Delta t \gtrsim 10$\,d), the observed optical
decay flattens, and in our last DEIMOS image the emission at the transient
location is clearly spatially resolved.  We interpret this to result from 
the emergence of an underlying host galaxy with $R \gtrsim 24.6$\,mag.
Neglecting the first point in the $R/r$-band light curve
(where the decay appears shallower), we find an optical decay
index of $\alpha_{\rm O}=1.02\pm0.02$ ($\chi^{2}=61.3$ for 10 degrees
of freedom).  The simultaneous GROND optical/near-infrared spectral-energy
distribution (SED) at $\Delta 
t=1.0$\,d is well fit by a power law with index $\beta_{\rm O}=0.63\pm0.10$ 
with no evidence for extinction in the host galaxy 
(Figure~\ref{fig:lcurve}).
All these results are broadly consistent with standard afterglow 
models (e.g., \citealt{gs02}) for expansion into a constant-density 
circumburst medium with electron index $p\approx2.5$ and a 
cooling break between the X-ray and optical bands (though close to the
X-rays, as the derived X-ray to optical spectral index at $\Delta t =1$\,d, 
$\beta_{\rm OX}=0.86$, is comparable to 
$\beta_{\rm X}=1.1^{+0.5}_{-0.3}$).  Furthermore, these properties are
typical of early X-ray (e.g., \citealt{ebp+09}) and optical
(e.g., \citealt{ckh+09}) afterglow light-curve behavior.

The temporal decay indices observed in the X-ray
and (especially) the optical are difficult to reconcile with 
post-jet-break evolution (e.g., \citealt{sph99}), as would be expected
for an off-axis orphan afterglow ($\alpha_{\mathrm{orphan}} \gtrsim 
2$)\footnote{Even if we allow the outburst time to vary freely in our 
power-law fits, the best-fit temporal indices in the X-ray and optical
are still $\lesssim2.0$.}.
Together with the rapid rise from our P48 nondetection 1.2\,hr before 
discovery, this further reinforces the association with GRB\,140226A, 
as it suggests iPTF14yb was initially viewed from within the jet opening 
angle.  The emergence of the host galaxy in the optical at $\Delta t 
\approx10$\,d greatly complicates our ability to detect any jet-break 
feature in the afterglow light curve, which would have offered robust 
support for such a geometry.

%%%%%%%%%%%%%%%%%%%%%%%%%%%%%%%%%%%%%%%%%%%%%%%%%%%%%%%%%%%%
\begin{figure*}[t!]
  \centerline{\includegraphics[width=16cm]{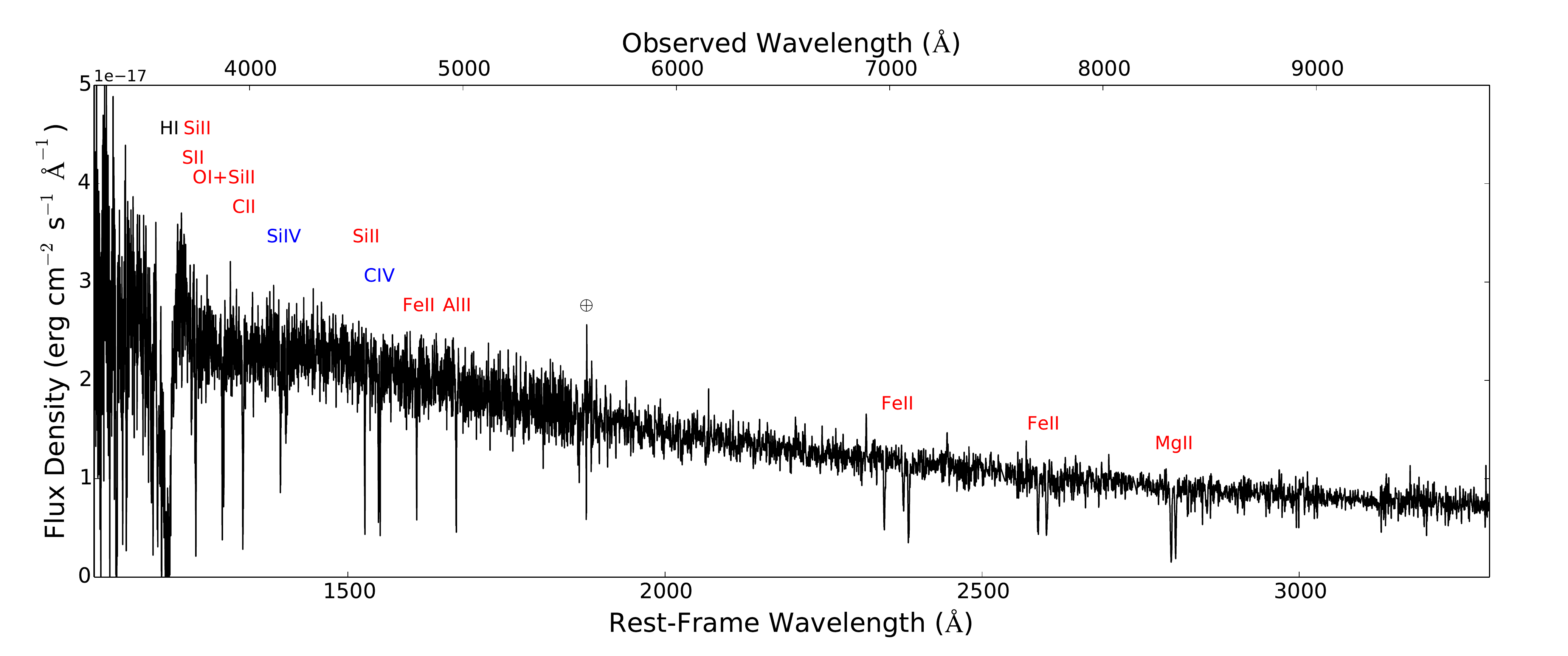}}
  \caption[]{%\small\it
  Keck/LRIS spectrum of iPTF14yb. Low-ionization (red labels) and
  high-ionization (blue labels) absorption lines from the host-galaxy
  interstellar medium, at a common redshift of 1.9733, are marked.}
  \label{fig:spec1}
\end{figure*}
%%%%%%%%%%%%%%%%%%%%%%%%%%%%%%%%%%%%%%%%%%%%%%%%%%%%%%%%%%%

Finally, the observed optical spectrum is typical of low-resolution 
spectra of the afterglows of long-duration GRBs (e.g., \citealt{fjp+09}), 
with DLA absorption and strong features from both low- and 
high-ionization-state metal transitions.  Unlike many other GRB afterglows, 
however, no fine-structure lines are apparent in the spectrum of iPTF14yb.  
This may be caused either by absorbing material that
is more distant from the explosion site, or simply a lack of 
sensitivity (in particular, spectral resolution).

%The detection of \ion{S}{2} can be used to estimate the metallicity
%of the host galaxy, as S does not deplete onto dust \citep{ss96}.
%With a measured rest-frame equivalent width of $W_{\mathrm{r}} = 0.54 \pm 
%0.10$\,\AA, we can calculate a lower limit to the column density 
%using the optically thin approximation:
%\begin{equation}
%\log_{10}(N_{\mathrm{X}}) = 1.23 \times 10^{20} 
%  \frac{W_{\mathrm{r}}}{\lambda_{\mathrm{r}}^2 f_{ij}}\,\mathrm{cm}^{-2}
%\end{equation}
%Here $\lambda_{\mathrm{r}}$ is the rest-frame wavelength of the transition,
%and $\lambda_{ij}$ is the oscillator strength.  For \ion{S}{2}, we
%calculate $\log_{10}(\mathrm{S\,II}) \gtrsim 15.6$, where the upper limit 
%results from the fact that the transition may be saturated at this low
%resolution \citep{p06}.  Even neglecting other ionization states,
%the \ion{S}{2} column density implies [S/H]$\equiv \log_{10}(\mathrm{S} /
%\mathrm{H}) - \log_{10}(\mathrm{S} / \mathrm{H})_{\odot} \gtrsim -0.2$.
%Similar limits are derived from the nondetection of \ion{Si}{2}
%$\lambda$ 1808.  While most long-duration GRBs occur in galaxies
%with sub-solar metallicities \citep{sgl09}, sight lines consistent
%with that observed for iPTF14yb are not entirely uncommon (e.g., 
%\citealt{srg+12,cfr+14}).

%===================================================
% Rates
%===================================================
\section{The Rate of Relativistic Transients}
\label{sec:rates}
Based on the results of the last two sections, we consider the association
between iPTF14yb and GRB\,140226A to be extremely robust.  This marks the
first time that a {\it bona fide} GRB afterglow (i.e., a cosmologically
distant, ultrarelativistic explosion) has been identified prior to
(and entirely independent of) a high-energy trigger.  While undoubtedly 
a technical achievement, we are more interested in what limits from the 
\textit{nondetection} of other, more exotic explosions may tell us 
about relativistic jet formation and collimation.  

Predictions for the rate of off-axis orphan-afterglow detection vary 
significantly depending on underlying model assumptions (e.g., average
opening angle, degree of lateral spreading of the ejecta).  For example,
at magnitude levels accessible to iPTF (peak $r \lesssim 20$), 
\citet{tp02} predicted that off-axis afterglows should outnumber on-axis 
GRBs by a factor of 3:1.  On the other hand, \citet{npg02} find that
on-axis GRBs will outnumber off-axis orphans for optical surveys with
a limiting magnitude shallower than $\sim23$.

In addition, while a relativistic explosion may be both jetted and viewed
on-axis, it may nonetheless lack prompt high-energy emission if the initial
Lorentz factor is not sufficiently high to overcome pair-production opacity.
Possible examples
include nearby relativistic supernovae such SN\,2009bb \citep{scp+10} and
SN\,2012ap \citep{csc+14}, as well as the fast-fading cosmological
transient PTF11agg \citep{ckh+13}.

To calculate the rate of relativistic transients, we assume here for 
simplicity that all events have light curves identical to that of iPTF14yb, 
and fold this through all P48 observations taken from
2013 Jan. 1 through 2014 March 1.  We only consider observations taken after
2013 Jan. 1 as we implemented significant improvements to our real-time
detection pipeline on that date.
We emphasize that our software requires at least two detections with
significance $> 5\sigma$ to filter out minor planets and image artifacts,
which does significantly reduce our sensitivity to such rapidly
fading transients (roughly by a factor of two).

Altogether, we find a total areal exposure of 
$A_{\mathrm{eff}}=24,637$\,deg$^{2}$\,d for iPTF14yb-like light curves.  
This implies an 
all-sky rate of relativistic transients of
\begin{align*}
\Re_{\mathrm{rel}} & \equiv \frac{N_{\mathrm{rel}}}{A_{\mathrm{eff}}} \\
 & =\frac{1}{24,637\,\mathrm{deg}^{2}\,\mathrm{d}}\times
\frac{365.25\,\mathrm{d}}{\mathrm{yr}}\times
\frac{41,253\,\mathrm{deg}^{2}}{\mathrm{sky}} \\
& = 610\,\mathrm{yr}^{-1}
\end{align*}
Assuming Poisson statistics, this implies a 68\% confidence 
interval of (110--2000)\,yr$^{-1}$.  We note that this value is actually
a lower limit, as it assumes we are 100\% efficient at \textit{discovering}
such sources in our data stream, even when they are significantly
\textit{detected} in our images.  For the remainder of this work we
assume our discovery efficiency, $\epsilon_{\mathrm{rel}}$, is $\sim 1$;
a more sophisticated analysis of the iPTF discovery efficiency suggests
this is a reasonable approximation (Urban et al., in prep.).

As a sanity check, we can compare this to the rate of on-axis long-duration
GRBs within the comoving volume out to $z\approx3$ (the approximate distance to
which P48 could detect iPTF14yb).  According to the BAT trigger 
simulations performed by \citet{lsg+14}, the all-sky rate of on-axis GRBs
out to $z=3$ is $\Re_{\mathrm{GRB}}=1455_{-112}^{+80}$\,yr$^{-1}$.  However, 
only a fraction of these events will have optical afterglows bright 
enough to detect by iPTF.  From unbiased samples of robotic 
follow-up observations of \textit{Swift} afterglows 
with moderate-aperture facilities (e.g., 
\citealt{ckh+09,gkk+11}), we infer that approximately 2/3 of 
long-duration GRBs have optical afterglows accessible to P48 
(peak $r\lesssim20$\,mag), or $\Re_{\mathrm{AG}}=970_{-74}^{+53}$\,yr$^{-1}$ 
GRBs. We conclude that our 
discovery rate is entirely consistent with the known population of 
on-axis \textit{Swift} events.

%%%%%%%%%%%%%%%%%%%%%%%%%%%%%%%%%%%%%%%%%%%%%%%%%%%%%%%%%%%%
\begin{figure*}[t!]
  \centerline{\includegraphics[width=16cm]{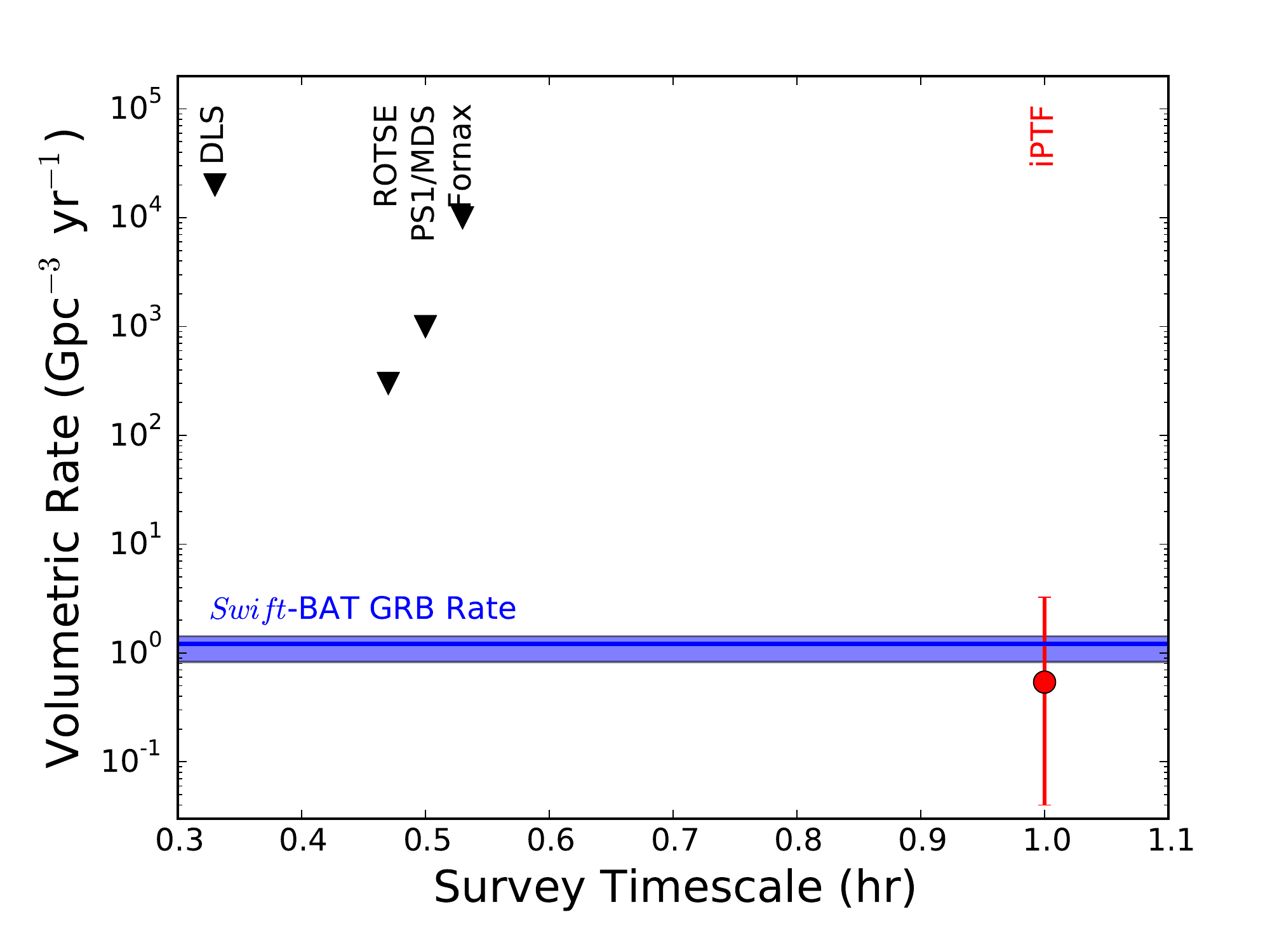}}
  \caption[]{%\small\it
  Rate of fast, luminous ($M\approx-27$\,mag) optical 
  transients, from iPTF and other surveys (\citealt{blc+13}, and
  references therein).  The all-sky 
  \textit{Swift} BAT GRB rate out to $z\approx3$
  is taken from \citet{lsg+14}.
  }
  \label{fig:rates}
\end{figure*}
%%%%%%%%%%%%%%%%%%%%%%%%%%%%%%%%%%%%%%%%%%%%%%%%%%%%%%%%%%%

Considering the comoving volume out to $z\approx3$, we can estimate
the volumetric rate of relativistic transients derived by iPTF to be
$\rho_{\mathrm{rel}}=0.54$\,Gpc$^{-3}$\,yr$^{-1}$.  
In addition to Poisson uncertainty in $\Re_{\mathrm{rel}}$, we 
include uncertainties on the value of $z_{\mathrm{max}}$: 
from 1.9733 (the redshift of iPTF14yb) to $\sim 6$ (where neutral 
H from the intergalactic medium precludes optical detection).  The 
resulting 68\% confidence interval on $\rho_{\mathrm{rel}}$ is thus 
(0.043--3.25)\,Gpc$^{-3}$\,yr$^{-1}$.
We plot this value in Figure~\ref{fig:rates} as a function of characteristic
survey time scale (e.g., cadence).  While iPTF repeats field visits on a wide
variety of time scales, we adopt here our typical internight cadence
of 1\,hr as representative for fast-fading sources.  Shown for comparison 
are limits for luminous ($M\approx-27$\,mag), rapidly fading optical 
transients from other surveys (\citealt{blc+13}, and references therein).  
These reported limits likely underestimate the sensitivity of these 
surveys to iPTF14yb-like transients, possibly by as much as an order of 
magnitude; for example, iPTF14yb would be detectable by PS1/MDS for 
approximately 1\,d, significantly longer than the 30\,min cadence.  
Nonetheless, by comparing with the all-sky rate of \textit{Swift} 
GRBs out to $z=3$ from \citet{lsg+14}, it is clear that iPTF is the first 
optical survey with sufficient sensitivity to detect on-axis GRBs 
independent of any high-energy trigger.    

Even with the relatively simple analysis performed here, our derived
limits appear to disfavor (though not entirely rule out) the most 
optimistic predictions for off-axis
orphan afterglows (e.g., \citealt{tp02}).  Furthermore, the rate of 
on-axis orphans (i.e., dirty fireballs) can also not be dramatically 
higher than the rate of long-duration GRBs, as was previously
argued based on the discovery of PTF11agg \citep{ckh+13}; see
\citet{ghv+00} for analogous limits in the X-rays.
A more detailed analysis would require, for example, more realistic models
of the afterglow luminosity function, redshift distribution, off-axis
emission, etc., enabling more robust limits to be placed on 
the typical GRB beaming angle and the optimal strategy for orphan 
searches with future wide-field optical surveys.  Such an analysis is
planned in a future work (Urban et al., in prep.).

Nonetheless, it is an exciting time in the search for orphan afterglows.
As several new wide-field optical transient surveys such as the Zwicky 
Transient Facility (ZTF) and the Large Synoptic Survey Telescope (LSST) 
prepare to see first light, the first {\it bona fide} detection is 
almost guaranteed to arrive in the coming years.  Future proposed wide-field
space missions such as ULTRASAT\footnote{See \url{http://www.weizmann.ac.il/astrophysics/ultrasat}.} would carry 
out sensitive searches for orphan afterglows as well.  Furthermore, new
wide-field radio surveys such as the Murchison Wide-Field Array 
(MWA), the Australian Square Kilometer Array Pathfinder (ASKAP),
and South African MeerKAT radio telescope, promise an even more
powerful census of relativistic explosions (though identifying them
may prove quite challenging, given their slow evolution at late
times).  iPTF14yb represents not only a technical milestone in 
fast-transient science, but also an important proof of concept for 
orphan-afterglow searches.

%%%%%%%%%%%%%%%%%%%%%%%%%%%%%%%%%%%%%%%%%%%%%%%%%%%%%%%%%%%%

%===================================================
% Acknowledgments
%===================================================
\acknowledgments
We thank David Jewitt for executing our Keck/LRIS ToO observations,
and Eran Ofek, Leo Singer, and Eric Bellm for comments on this 
manuscript.  A.L.U.~was supported by NSF grants 
PHY-0970074 and PHY-1307429 at the UWM Research Growth Initiative.  
J.F.G.~acknowledges the Sofja Kovalevskaja award to 
P.~Schady from the Alexander von Humboldt Foundation Germany.  
D.A.K~thanks TLS Tautenburg for financial support.   
The work of A.V.F.~was made possible by NSF grant AST-1211916, the TABASGO 
Foundation, Gary and Cynthia Bengier, and the Christopher R. Redlich Fund.
J.X.P.~received funding from NASA grants NNX13AP036 and NNX14AI95G.

This paper is based in part on observations obtained with the P48 
Oschin telescope as part of the Intermediate Palomar Transient 
Factory project, a scientific collaboration among the Caltech, 
LANL, UW-Milwaukee, the Oskar Klein Center, the Weizmann Institute of 
Science, the TANGO Program of the University System of Taiwan, and 
the Kavli IPMU.  LANL participation in iPTF is supported by the US 
Department of Energy as part of the Laboratory of Directed Research and 
Development program.  The National Energy 
Research Scientific Computing Center provided staff, computational 
resources, and data storage for this project.  Part of the funding for GROND (both 
hardware and personnel) was generously granted from the 
Leibniz-Prize to Prof. G. Hasinger (DFG grant HA 1850/28-1). 
Some of the data presented herein were obtained at the W.~M. Keck
Observatory, which is operated as a scientific partnership among the
California Institute of Technology, the University of California, and
NASA; the observatory was made possible by the generous financial
support of the W.~M. Keck Foundation.

We thank the RATIR project team and the staff of the Observatorio 
Astron\'{o}mico Nacional on Sierra San Pedro M\`{a}rtir. RATIR is a 
collaboration between the University of California, the Universidad 
Nacional Auton\'{o}ma de M\'{e}xico, NASA Goddard Space Flight Center, 
and Arizona State University, benefiting from the loan of an H2RG 
detector and hardware and software support from Teledyne Scientific 
and Imaging. RATIR, the automation of the Harold L. Johnson Telescope 
of the Observatorio Astron\'{o}mico Nacional on Sierra San Pedro 
M\'{a}rtir, and the operation of both are funded through NASA grants 
NNX09AH71G, NNX09AT02G, NNX10AI27G, and NNX12AE66G, CONACyT grants 
INFR-2009-01-122785 and CB-2008-101958, UNAM PAPIIT grant IN113810, 
and UC MEXUS-CONACyT grant CN 09-283.

%===================================================
% Facility Keywords
%===================================================
\vspace{0.5cm}

{\it Facilities:} \facility{PO: 1.2m (PTF)}, \facility{Keck:I (LRIS)},
\facility{Magellan: Baade (IMACS)}, \facility{Lulin (TRIPOL)},
\facility{Max Planck:2.2m (GROND)}, \facility{OANSPM:HJT (RATIR)},
\facility{VLA}, \facility{CARMA}, \facility{Swift (XRT)}

%===================================================
% Bibliography
%===================================================
\bibliographystyle{apj}
%\bibliography{master}

%===================================================
% Tables
%===================================================
\clearpage
\begin{deluxetable*}{lcccc}
\tabletypesize{\scriptsize}
\tablecaption{Optical/Near-Infrared Observations of iPTF14yb}
\tablewidth{0pt}
\tablehead{
\colhead{Date} & \colhead{Telescope/Instrument} & \colhead{Filter} &
\colhead{Exposure Time} & \colhead{Magnitude\tablenotemark{a}} \\
\colhead{(MJD)} & & & (s) &
}
\startdata
56714.379 & P48/CFHT12k & $r$ & 60.0 & $> 21.16$ \\
56714.429 & P48/CFHT12k & $r$ & 60.0 & $18.16 \pm 0.03$ \\    
56714.502 & P48/CFHT12k & $r$ & 60.0 & $19.77 \pm 0.07$ \\
56714.548 & P48/CFHT12k & $r$ & 60.0 & $20.24 \pm 0.08$ \\ 
56714.723 & Lulin/TRIPOL & $g^{\prime}$ & 2400.0 & $21.17 \pm 0.13$ \\
56714.723 & Lulin/TRIPOL & $r^{\prime}$ & 2400.0 & $21.07 \pm 0.11$ \\
56714.723 & Lulin/TRIPOL & $i^{\prime}$ & 2400.0 & $20.79 \pm 0.11$ \\
56714.835 & Lulin/TRIPOL & $g^{\prime}$ & 2700.0 & $21.61 \pm 0.18$ \\
56714.835 & Lulin/TRIPOL & $r^{\prime}$ & 2700.0 & $21.28 \pm 0.13$ \\
56714.835 & Lulin/TRIPOL & $i^{\prime}$ & 2700.0 & $21.00 \pm 0.11$ \\
56715.379 & La Silla/GROND & $g^{\prime}$ & 869.9 & $22.32 \pm 0.04$ \\
56715.379 & La Silla/GROND & $r^{\prime}$ & 869.9 & $22.07 \pm 0.03$ \\
56715.379 & La Silla/GROND & $i^{\prime}$ & 869.9 & $22.04 \pm 0.06$ \\
56715.379 & La Silla/GROND & $z^{\prime}$ & 869.9 & $21.82 \pm 0.08$ \\
56715.379 & La Silla/GROND & $J$ & 869.9 & $> 21.6$ \\    
56715.379 & La Silla/GROND & $H$ & 869.9 & $> 21.0$ \\
56715.379 & La Silla/GROND & $K$ & 869.9 & $> 19.8$ \\
56715.413 & SPM/RATIR & $r^{\prime}$ & 14076.0 & $22.24 \pm 0.09$ \\
56715.413 & SPM/RATIR & $i^{\prime}$ & 14076.0 & $21.99 \pm 0.09$ \\
56715.413 & SPM/RATIR & $z^{\prime}$ & 5904.0 & $21.96 \pm 0.24$ \\
56715.413 & SPM/RATIR & $Y$ & 5904.0 & $22.10 \pm 0.35$ \\
56715.413 & SPM/RATIR & $J$ & 5904.0 & $21.87 \pm 0.35$ \\
56715.413 & SPM/RATIR & $H$ & 5904.0 & $> 21.58$ \\
56716.362 & La Silla/GROND & $g^{\prime}$ & 1871.3 & $23.15 \pm 0.06$ \\
56716.362 & La Silla/GROND & $r^{\prime}$ & 1871.3 & $23.04 \pm 0.06$ \\
56716.362 & La Silla/GROND & $i^{\prime}$ & 1871.3 & $22.82 \pm 0.11$ \\
56716.362 & La Silla/GROND & $z^{\prime}$ & 1871.3 & $22.66 \pm 0.11$ \\
56716.362 & La Silla/GROND & $J$ & 1871.3 & $> 21.5$ \\    
56716.362 & La Silla/GROND & $H$ & 1871.3 & $> 20.9$ \\
56716.362 & La Silla/GROND & $K$ & 1871.3 & $> 19.8$ \\
56716.435 & SPM/RATIR & $r^{\prime}$ & 9468.0 & $23.24 \pm 0.32$ \\
56716.435 & SPM/RATIR & $i^{\prime}$ & 9468.0 & $22.48 \pm 0.20$ \\
56716.435 & SPM/RATIR & $z^{\prime}$ & 4032.0 & $> 22.09$ \\
56716.435 & SPM/RATIR & $Y$ & 4032.0 & $> 21.72$ \\
56716.656 & Keck-I/LRIS & $g^{\prime}$ & 540.0 & $23.21 \pm 0.04$ \\
56716.656 & Keck-I/LRIS & $R$ & 480.0 & $23.13 \pm 0.05$ \\
56717.641 & Keck-I/LRIS & $g^{\prime}$ & 720.0 & $23.68 \pm 0.02$ \\
56717.642 & Keck-I/LRIS & $R$ & 640.0 & $23.51 \pm 0.03$ \\
56725.373 & Baade/IMACS & $R$ & 1500.0 & $24.70 \pm 0.19$ \\
56742.557 & Keck-II/DEIMOS & $R$ & 2100.0 & $24.64 \pm 0.09$ \\
\enddata
\tablenotetext{a}{Reported magnitudes are in the AB system and have been
corrected for a foreground Galactic extinction of $E(B-V)=0.016$\,mag 
\citep{sf11}.}
\label{tab:photometry}
\end{deluxetable*}

%%%%%%%%%%%%%%%%%%%%%%%%%%%%%%%%%%%%%%%%%%%%%%%%%%%%%%%%%%%%

%===================================================
% The End
%===================================================
\end{document}